\documentclass[a4paper]{article}
\usepackage{epsfig,amsmath,amsfonts,amssymb,setspace,multirow,textcomp}
\usepackage[T1]{fontenc}
\makeatletter
\renewcommand{\@oddhead}{\textit{Advances in Astronomy and Space Physics} \hfil}
\renewcommand{\@evenfoot}{\hfil \thepage \hfil}
\renewcommand{\@oddfoot}{\hfil \thepage \hfil}
\makeatother

\renewenvironment{thebibliography}[1]{\begin{oldthebibliography}{#1}\setlength{\parskip}{0ex}\setlength{\itemsep}{0ex}}{\end{oldthebibliography}}
\begin{document}
\fontsize{11}{11}\selectfont % the font size cannot be changed in any case!
%  insert your title, authors information and text instead of the one provided below
\title{Cosmic-Ray Extremely Distributed Observatory: a global cosmic ray detection framework}
\author{\textsl{O.~Sushchov$^{1}$, P.~Homola$^{1}$, N.~Dhital$^{1}$, \L{}.~Bratek$^{1}$, P.~Pozna\'nski$^{1, 2}$,}\\
\textsl{T.~Wibig$^{3, 4}$, J.~Zamora-Saa$^{5}$, K.~Almeida~Cheminant$^{1}$, D.~Alvarez~Castillo$^{5}$, D.~G\'ora$^{1}$,}\\
\textsl{P.~Jagoda$^{6, 1}$, J.~Ja\l{}ocha$^{7}$, J.\,F.~Jarvis$^{1, 8}$, M.~Kasztelan$^{9}$, K.~Kopa\'nski$^{1}$, M.~Krupi\'nski$^{1}$,}\\
\textsl{M.~Micha\l{}ek$^{2, 1}$, V.~Nazari$^{1, 4}$, K.~Smelcerz$^{2, 1}$, K.~Smolek$^{10}$, J.~Stasielak$^{1}$ and M.~Su\l{}ek$^{2, 1}$}}
\date{\vspace*{-6ex}}
\maketitle
\begin{center} {\small $^{1}$Institute of Nuclear Physics Polish Academy of Sciences, Radzikowskiego 152, 31-342 Cracow, Poland\\
$^{2}$Cracow University of Technology, Warszawska 24, 31-155 Cracow, Poland\\
$^{3}$Physics Education Lab, Faculty of Physics and Applied Informatics, University of \L{}\'od\'z,\\
149/158 Pomorska str., 90-236 \L{}\'od\'z, Poland\\
$^{4}$Cosmic Ray Laboratory, Astrophysics Division, National Centre for Nuclear Research,\\
69 Pu\l{}ku Strzelc\'ow Kaniowskich Str., 90-558 \L{}\'od\'z, Poland\\
$^{5}$Joint Institute for Nuclear Research, 141980 Dubna, Russia\\
$^{6}$AGH University of Science and Technology, al.~Mickiewicza 30, 30-059 Cracow, Poland\\
$^{7}$Institute of Physics, Cracow University of Technology, PL-30084 Cracow, Poland\\
$^{8}$School of Physical Sciences, Open University, Buckinghamshire, MK7 6AA, United Kingdom\\
$^{9}$National Centre for Nuclear Research, Andrzeja So\l{}tana 7, 05-400 Otwock-\'Swierk, Poland\\
$^{10}$Institute of Experimental and Applied Physics, Czech Technical University in Prague,\\
Horsk\'a 3a/22, 12800, Praha2, Czech Republic\\
{\tt oleksandr.sushchov@ifj.edu.pl}}\\
\end{center}

\begin{abstract}
The main objective of the Cosmic-Ray Extremely Distributed Observatory (CREDO) is the detection and analysis of extended cosmic ray phenomena, so-called super-preshowers (SPS), using existing as well as new infrastructure (cosmic-ray observatories, educational detectors, single detectors etc.). The search for ensembles of cosmic ray events initiated by SPS is yet an untouched ground, in contrast to the current state-of-the-art analysis, which is focused on the detection of single cosmic ray events. Theoretical explanation of SPS could be given either within classical (e.g., photon-photon interaction) or exotic (e.g., Super Heavy Dark Matter decay or annihilation) scenarios, thus detection of SPS would provide a better understanding of particle physics, high energy astrophysics and cosmology. The ensembles of cosmic rays can be classified based on the spatial and temporal extent of particles constituting the ensemble. Some classes of SPS are predicted to have huge spatial distribution, a unique signature detectable only with a facility of the global size. Since development and commissioning of a completely new facility with such requirements is economically unwarranted and time-consuming, the global analysis goals are achievable when all types of existing detectors are merged into a worldwide network. The idea to use the instruments in operation is based on a novel trigger algorithm: in parallel to looking for neighbour surface detectors receiving the signal simultaneously, one should also look for spatially isolated stations clustered in a small time window. On the other hand, CREDO strategy is also aimed at an active engagement of a large number of participants, who will contribute to the project by using common electronic devices (e.g., smartphones), capable of detecting cosmic rays. It will help not only in expanding the geographical spread of CREDO, but also in managing a large manpower necessary for a more efficient crowd-sourced pattern recognition scheme to identify and classify SPS. A worldwide network of cosmic-ray detectors could not only become a unique tool to study fundamental physics, it will also provide a number of other opportunities, including space-weather or geophysics studies. Among the latter one has to list the potential to predict earthquakes by monitoring the rate of low energy cosmic-ray events. The diversity of goals motivates us to advertise this concept across the astroparticle physics community.\\[1ex]
{\bf Key words:} dark matter, cosmic rays, data analysis
\end{abstract}

\section*{\sc introduction}
\begin{figure}[!h]
\centering
\begin{minipage}[t]{.45\linewidth}
\centering
\epsfig{file = 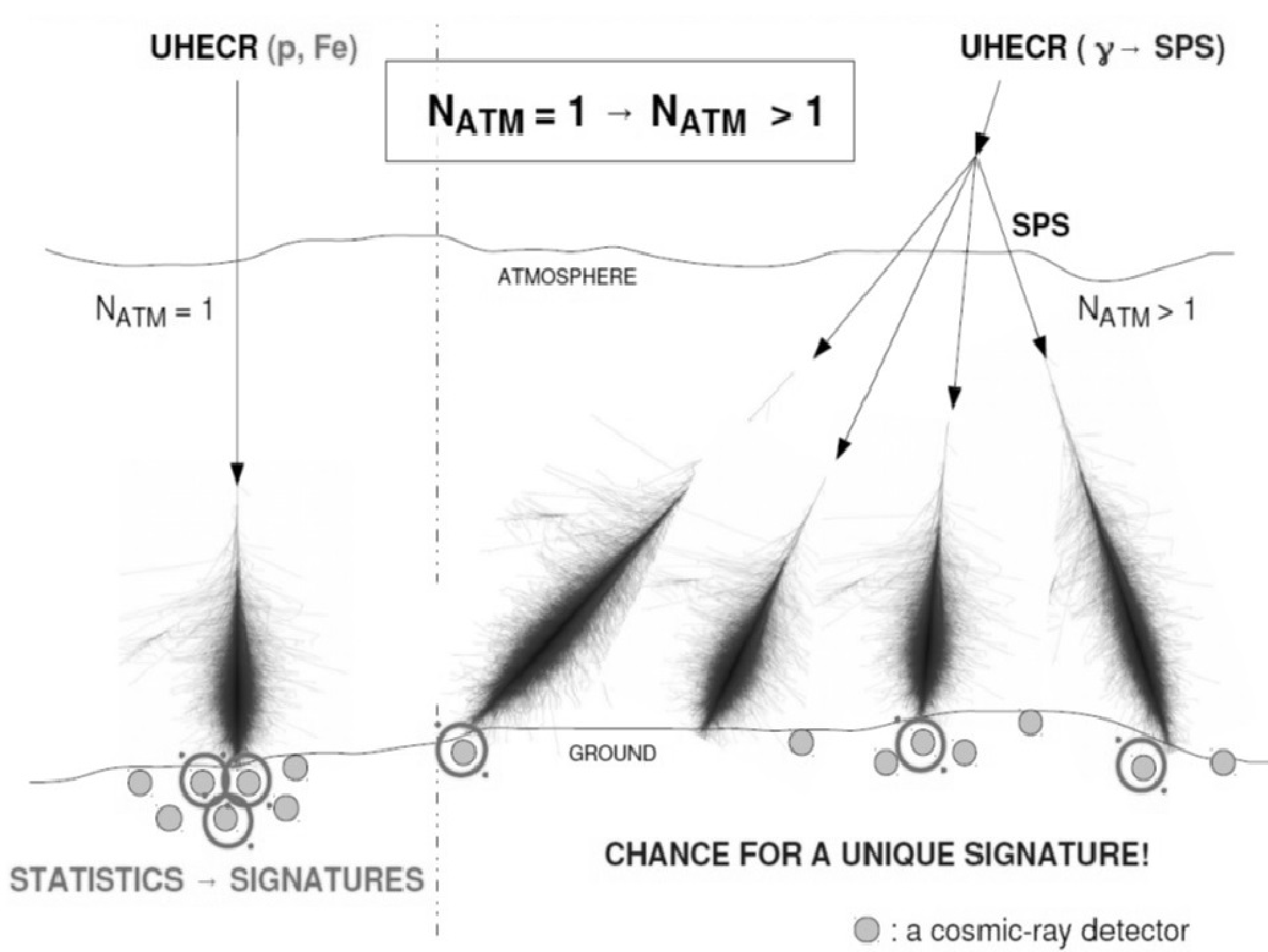,width = .85\linewidth}
\caption{Approaches to cosmic-ray research: state-of-the-art (left) and proposed by CREDO (right).}\label{fig1}
\end{minipage}
\hfill
\begin{minipage}[t]{.45\linewidth}
\centering
\epsfig{file = 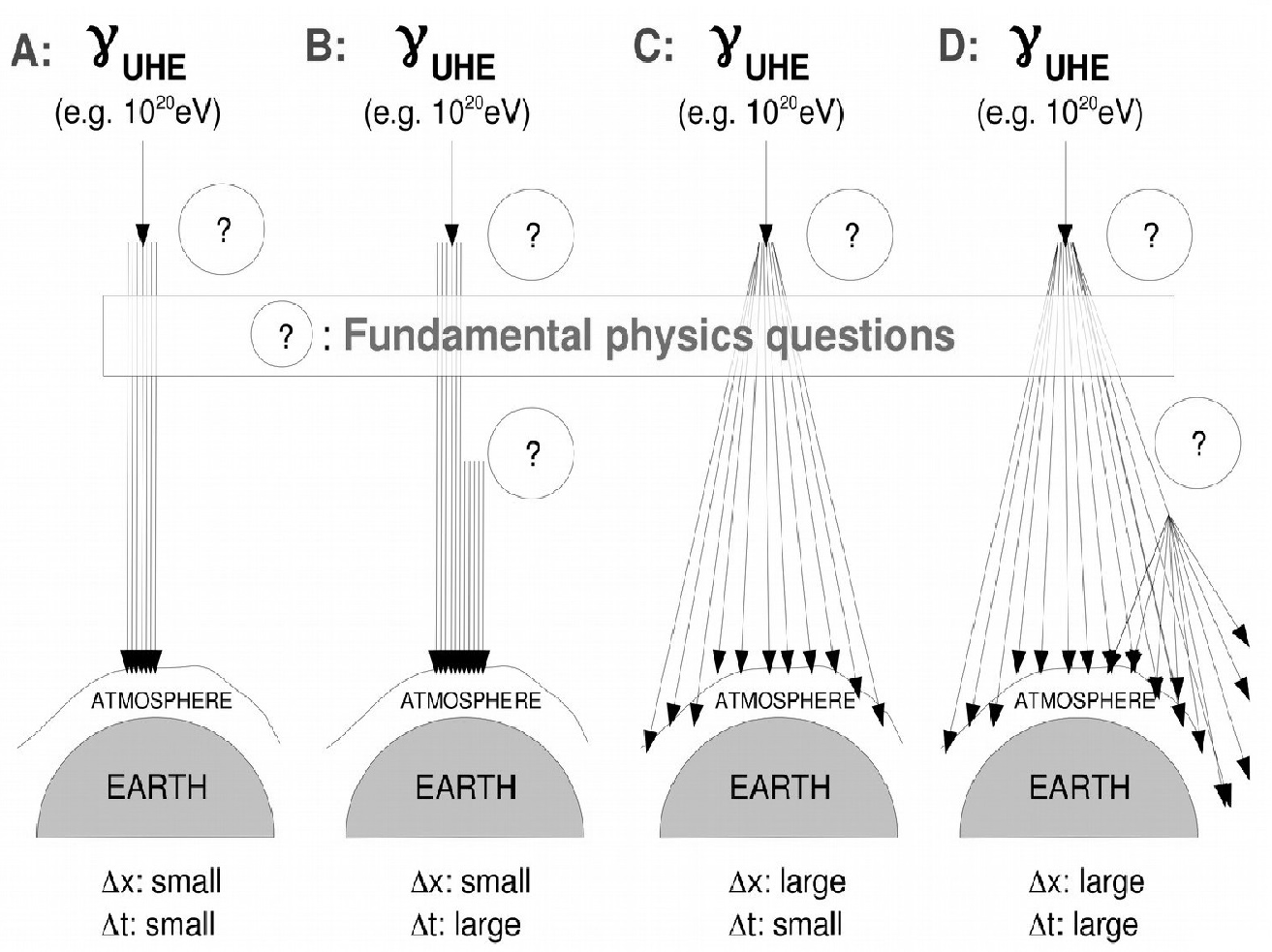,width = .85\linewidth}
\caption{SPS classification. Question marks denote unknown/uncertain physics processes at EHE.}\label{fig2}
\end{minipage}
\end{figure}

\indent \indent At least two fundamental problems of modern astrophysics are addressed by the CREDO project: i) the
nature of Dark Matter and ii) explanation of the existence of cosmic rays with energies exceeding ${10}^{20}$~eV
(hereafter referred to as extremely-high energy cosmic rays - EHECR). The hypothesis to be tested and verified is 
that two above mentioned puzzles of nature could be explained with just one scenario, known in the literature as Super
Heavy Dark Matter (SHDM) decay or annihilation (see e.g. \cite{Chung99}). It assumes production of supermassive (i.e.
$M \geq {10}^{23}$ eV) particles in the early Universe, after the inflation phase. Such particles could annihilate or
decay presently, leading to the production of jets containing mainly photons. The energies of these photons
could easily be of the order of ${10}^{20}$ eV, the value that seems to be out of reach in the acceleration processes
in the potential sources. The key prediction of the scenarios in the SHDM group is that the EHECR flux
observed at the Earth should be dominated by photons (see e.g. \cite{Rubtsov06}). On the other hand, the highest energy
events observed by the leading collaborations, Pierre Auger Observatory and Telescope Array, are not considered photon
candidates if the present state-of-art air shower reconstruction procedures are applied. In fact, there are no
photon candidates also in the data below ${10}^{20}$ eV, collected by these and other observatories, leading to the
very stringent upper limits (\cite{Bleve15}, \cite{Abu13}). There are several theoretical models trying to explain the non-observation of extremely-high energy (EHE) photons at Earth. For instance, Lorentz Invariance Violation (see e.g. \cite{Jacobson06}) scenarios predict that the lifetime of such a photon could be very short ($<$$<$1 sec), thus such photons would have negligible chances to reach the Earth. The main question to be asked is about the assumptions under which the upper limits can be interpreted as constraints to SHDM or other exotic scenarios - as it is commonly concluded. Generally, there are at least two main doubts about such conclusions: the present state-of-the-art analysis does not take into account mechanisms that could lead i) to a good mimicking of hadronic air showers with the showers induced by photons, and ii) to the efficient screening/cascading of EHE photons on their way to the Earth, so that the products of such screening/cascading are out of reach of the presently operating observatories, which is interpreted as non-observation of EHE photons. If the doubt i) is founded in reality, we do have photons in the data but we can not identify them properly. If the doubt ii) addresses the real properties of cosmic rays, then we have no chance to see most of the photons that travel towards us. Both doubts obviously question the conclusion about constraining the SHDM scenarios by the presently accepted upper limits to photon flux. Such a conclusion can only be accepted with the assumption that both doubts are irrelevant. The key point of CREDO is that this assumption can be experimentally tested with the use of available infrastructure, technique and analysis methods in a novel way. Different approach could be applied to observe the results of a decay of a EHE photon, which is assumed to be a very extended cascade consisting mainly of photons with different energies. We consider these cascades in a general way, no matter by what process they are initiated, and call them super-preshowers. More detailed explanation of this notion could be found in the following paragraph, but here we only present the difference between the existing technique and the proposed one, as shown on Fig.~\ref{fig1}. The state-of-the-art approach is being focused on uncorrelated in time cosmic rays arriving at the top of the atmosphere (${N}_{ATM}$=1), while the novel one is aimed at the detection of cosmic-ray cascades: a group of cosmic rays correlated in time (${N}_{ATM}$>1). The concept that EHE events could be registered not as a single event but rather as bursts of events of lower energy was already discussed (see, e.g., \cite{cerncourier} and references therein). It was proposed that very-large-area networks of cosmic-ray detectors were needed to search for large area cosmic-ray coincidences and their sources, giving at the same time the opportunity to involve high school teachers and scientists in the excitement of fundamental research. Some of them have been operating since late 1990-s (e.g., QuarkNet \cite{QN13}, HiSPARC \cite{HiSPARC05}, The Roland Maze Project \cite{Maze06}, CZELTA \cite{czelta11}, Shower of Knowledge \cite{Dubna}, etc.). Within CREDO, the use of all existing detectors with addition of the mobile ones ensures wider geographical spread and engagement of larger number of participants, increasing all in all observational capacity. In addition, since the research to be undertaken is on the edge of unknown, one cannot be surprised if a completely unexpected fundamental discovery will be made - something beyond the state-of-the-art models. In this regard, the involvement of professional detectors by CREDO would strengthen the search for signatures of new physics scenarios.

\section*{\sc scientific goals}

\indent \indent SHDM scenarios, mentioned above, might be directly connected to another important astrophysical puzzle~- 
existence of EHECRs. The problem is wider than the existence of particles with energies exceeding  ${10}^{20}$~eV itself. 
Not explained so far is also the fact that the reconstructed trajectories of EHECR propagation do not show clear correlation with potential astrophysical sources \cite{Abreu11}. There are also discrepancies in the major experiments' interpretations of the observed energy spectrum cutoff (\cite{Schulz15}, \cite{Abu15}), as well as some general problems in the multi-channel data explanation \cite{Aartsen16}. Numerous and yet inconclusive efforts of solving the EHECR mystery with the current approach could be regarded as an argument that alternative scenarios capable of EHE particles' generation without an absolute need for correlation with the sources should be taken into serious consideration as well. While gravitational properties of SHDM particles are rather unquestioned, their distribution seems to be disputable, e.g. not every galaxy can be an SHDM source. It is, for instance, not clear in the case of our Galaxy \cite{Sikora15}. SHDM sources could hardly be registered in the case when SHDM distribution is nearly uniform on a scale of super galaxy cluster, or if EHE photons have smaller horizon of propagation than it is believed. The other point is that constraints on the sources or processes leading to the production of EHE photons could be inferred considering several general assumptions, concerning electrodynamics at energies in the Grand Unified Theory (GUT) scale \cite{Maccione08}, hadronic properties of EHE photons \cite{Levy}, or space-time structure \cite{Ellis01}. Modelling EHE photons' propagation through space requires assumptions, which are difficult to verify. Besides, physics processes at GUT energies can hardly be predicted, moreover, deviations from Standard Models in cosmology and particle physics are expected because of the obvious disagreement between General Relativity and Quantum Field Theory (see e.g. \cite{Almheiri12}). Thus, any conclusions considering GUT uncertainties, in particular, the ones based on the constraints to EHE photon flux, are disputable. To put the main idea in the simplest way one considers a hypothetic mechanism leading to a cascading of most of the EHE photons before they reach Earth, leading to the efficient shrinking of their astrophysical horizon. If such a mechanism or process occurs in reality, EHE photons have very small chances to reach the Earth. But one can expect that resulting particles, arriving most likely as vast electromagnetic cascades, could be observed at the Earth. An example of such a process is the so-called preshower effect (see, e.g., \cite{Erber66}, \cite{McBreen81}, \cite{Homola05} etc.), occurring as a result of an interaction of a EHE photon and secondary electrons with geomagnetic field.  The word ``preshower'' is used to describe the result of the initial interaction (many particles instead of a single primary EHE photon), and emphasize the location of the interaction vertex (above the atmosphere, i.e. before the extensive air showers (EAS) are initiated by the preshower particles). To generalize this notion, the ``super-preshower'' (SPS) term was introduced to define a cascade of electromagnetic particles originated above the Earth atmosphere, regardless of the initiating process and distance from the Earth. SPS can be classified with respect to their principal observable properties: spread in space  ($\Delta{x}$) and time ($\Delta{t}$). A schematic of such classification is shown in Fig.~\ref{fig2} \cite{Homola16}. A cascade initiated by the interaction of a EHE photon with the geomagnetic field is contained within a few square centimeters (cases A and B) \cite{Homola05}. For the same effect occurring at the vicinity of the Sun, one would expect particles with narrow time distribution but large spatial extent as the cascade arrives at the top of the Earth's atmosphere (class ``C'') \cite{Dhital17}. Besides, less known scenarios could combine quite significant distribution of the particles' arrival times with narrow spatial distribution (class ``B''), or both distributions are extended (class ``D''). Thus, the worldwide network of cosmic ray detectors, planned to be organized within CREDO, will become a unique facility to study fundamental physics laws. Moreover, it could be also used in interdisciplinary research, for instance, geophysics and space-weather studies. Among the yet unprobed geophysics research one could name capability of CREDO framework to provide studies of correlation between the rate of low energy cosmic rays secondaries, usually regarded as a background signal, and seismic effects. Existence of such correlations is being discussed for quite a long time (\cite{Morozova00}, \cite{Antonova09}, \cite{Kovalyov14}) and could be studied to provide major earthquakes' forecast. Widely distributed infrastructure might increase the chances to verify the hypothesis, which, in turn, could help saving human lives in seismically dangerous regions.

\section*{\sc detectors}

\indent \indent State-of-the-art cosmic-ray research is focused on detecting single air showers, and, as was explained above, it is
sensitive to super-preshowers type A and B, but not capable to register SPS type C or D. If an observatory is exposed only to SPS particles of energies below the operating trigger, there will be no observation. On the other hand, one can imagine an alternative detection method of SPS-C and SPS-D with already existing infrastructure. If cosmic-ray detectors of different types (professional cosmic-ray observatories, educational networks, university particle counters and smartphones, equipped with photo-sensors and appropriate applications) were organized in a global network, it could become a powerful tool aimed at search for ensembles of cosmic particles arriving at the Earth in huge fronts (up to the Earth size). Therefore, the main goal of CREDO is detection of very extended cascades in a global scale, when secondary particles have large scale time correlations. For obvious reasons (financial, time-consuming), the development and commissioning of a new facility serving these goals is infeasible, and the only way to achieve this objective seems to be to organize a single, global framework mostly composed of the existing detectors covering area at best of global scale. Thus, CREDO could easily combine i) existing cosmic-ray detectors, both educational and state-of-the-art facilities, ii) new detectors, which can be designed and commissioned easily and economically, and iii) common electronic devices with photo-sensors, capable of cosmic ray detection
using specially designed applications. The data registered by interconnected detectors (via a common network), could be stored in a single database, and used for a real-time or a near real-time monitoring and data analysis. For the time being, CREDO is able to perform analysis of limited amount of real cosmic ray data from the projects where at least part of the data is public. In parallel CREDO approaches the cosmic-ray related projects, which restrict access to their data to discuss policies for possible exchanges and cooperation. In addition, CREDO will have new cosmic ray detectors within its framework. Equipping various cities all over the globe with such detectors is of great importance, because it will shed light on CREDO overall performance, and will likely provide triggering or alerting nearby more basic but numerous detectors when needed. Although it would be ideal to have as many detectors as possible to realize the CREDO strategy aimed at new physics/exotic searches, number of such detectors which can be realistically installed is limited by the financial constraints. Also, various scenarios to be tested within CREDO will likely have different requirements in terms of the number of detectors, which will be estimated for each scenario separately. Electronic devices like smartphones, equipped with cameras capable to detect cosmic rays, are planned to become another important component of CREDO. Cosmic rays detection with smartphones has already been probed by two collaborations: 
the Distributed Electronic Cosmic-ray Observatory (DECO) \cite{DECO}, and Cosmic RAYs Found In Smartphones (CRAYFIS) \cite{CRAYFIS}. Additionally, a new smartphone application for cosmic ray detection is being developed within CREDO. It is conceived to serve as an open source project for users and developers across the world. Participation by a large number of enthusiasts, non-professional scientists, also known as ``citizen science'', is another key idea of CREDO, aimed at scientific (geographical expansion of CREDO and thus help to reach its main goals), educational goals as well as popularization of science, stimulating curiosity and development of creative way of thinking among young people all over the world. Given a wide range of objectives, including education, fundamental science goals and a possibility to perform interdisciplinary studies in areas like space-weather and earthquake prediction, participation from a large number of science enthusiasts is expected and warmly welcomed. Once the application is fully developed and is in use, all relevant information regarding cosmic ray hits, locations of the detectors and the timestamps of the hits are sent to CREDO computing infrastructure for storage and analysis.

\section*{\sc analysis}

\begin{figure}[!h]
\centering
\begin{minipage}[t]{.45\linewidth}
\centering
\epsfig{file = 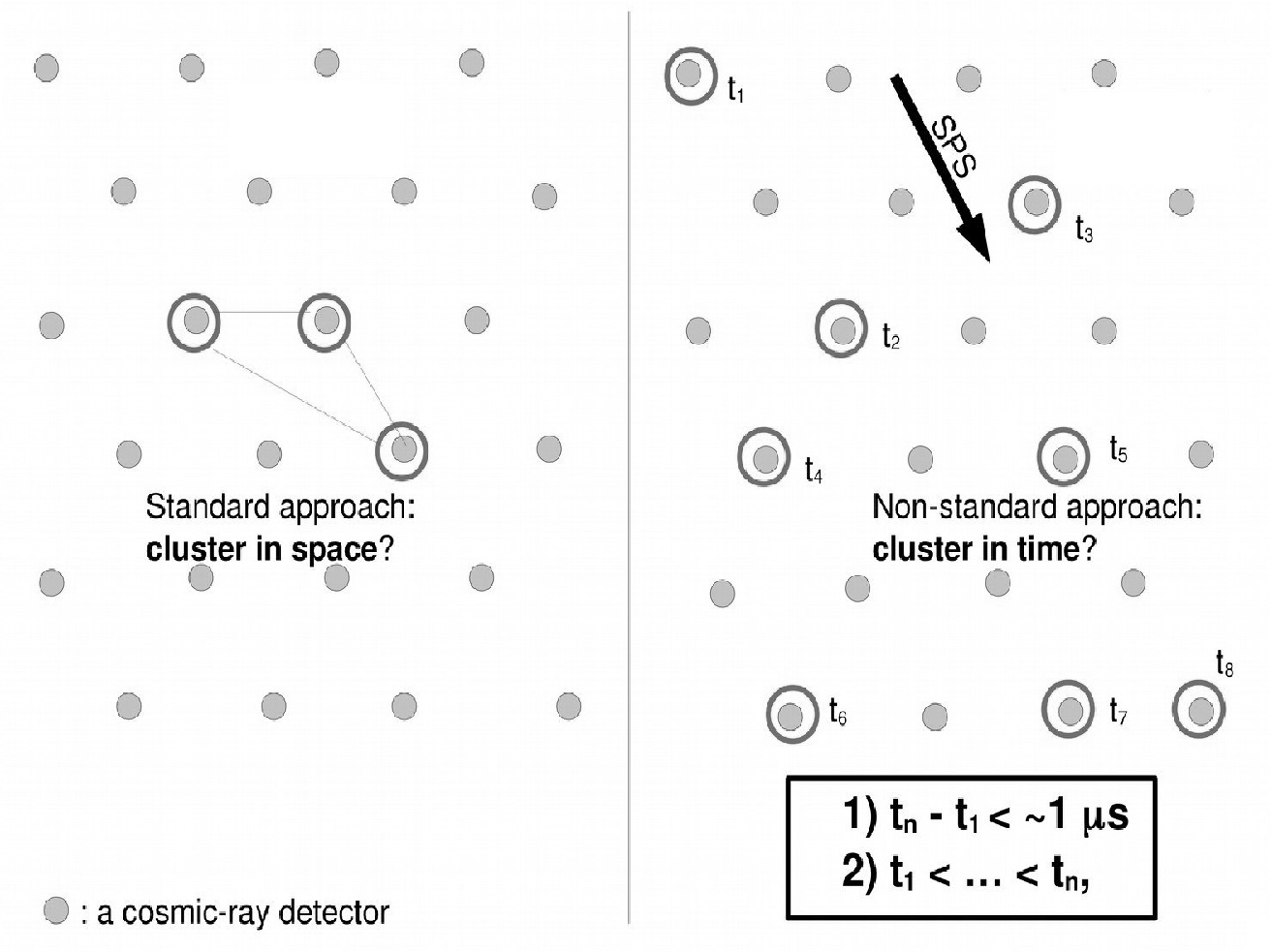,width = .85\linewidth}
\caption{SPS searching strategy schemes: standard (left) and proposed (right).}\label{fig3}
\end{minipage}
\end{figure}

\indent \indent The easiest and most obvious way of identifying a super-preshower is to analyze the already available data. 
It might be very difficult in the case of type A and B SPS. As it was already explained, if SPS-initiated air showers are present in the available data, the currently used set of observables might not be enough to identify SPS properly, if the SPS A/B primaries can
mimic the nuclei well. In the case of SPS C/D the perspective is more optimistic for large instruments like the Pierre Auger
Observatory. Let us consider an SPS C/D with most of the particles below the event trigger level of a big observatory.
In this case, the low energy secondaries, arriving at the surface array, are registered if there is no event-level trigger at the same time. These particles are considered a background and the corresponding data are processed only within a working buffer, and are not stored later if no standard event is found. This ignored background could be treated as a signal within the SPS C/D strategy. Unlike the standard trigger algorithm, developed to search for a group of neighbour stations receiving the signal simultaneously, one might want to look for a group of spatially isolated stations receiving the signal within the same or a similar time window. This analysis/trigger idea is illustrated in fig.~\ref{fig3} \cite{Homola16}. Within this approach the number of isolated stations, required for a clearly non-random signal, should be studied using Monte-Carlo simulations of SPS, subsequent air showers and detector response. This strategy could be implemented in a straightforward way in the global infrastructure like CREDO. Additional difficulties will be met while optimizing the details: temporal and geographical windows, studying various arrival directions and front geometries of SPS, considering possible delays in time, etc., but the key analysis idea should remain unchanged: search for clearly non-random signal arrival time patterns extending over an area larger than in case of a single extensive air shower.

\section*{\sc credo monitor}

\begin{figure}[!h]
\centering
\begin{minipage}[t]{.45\linewidth}
\centering
\epsfig{file = 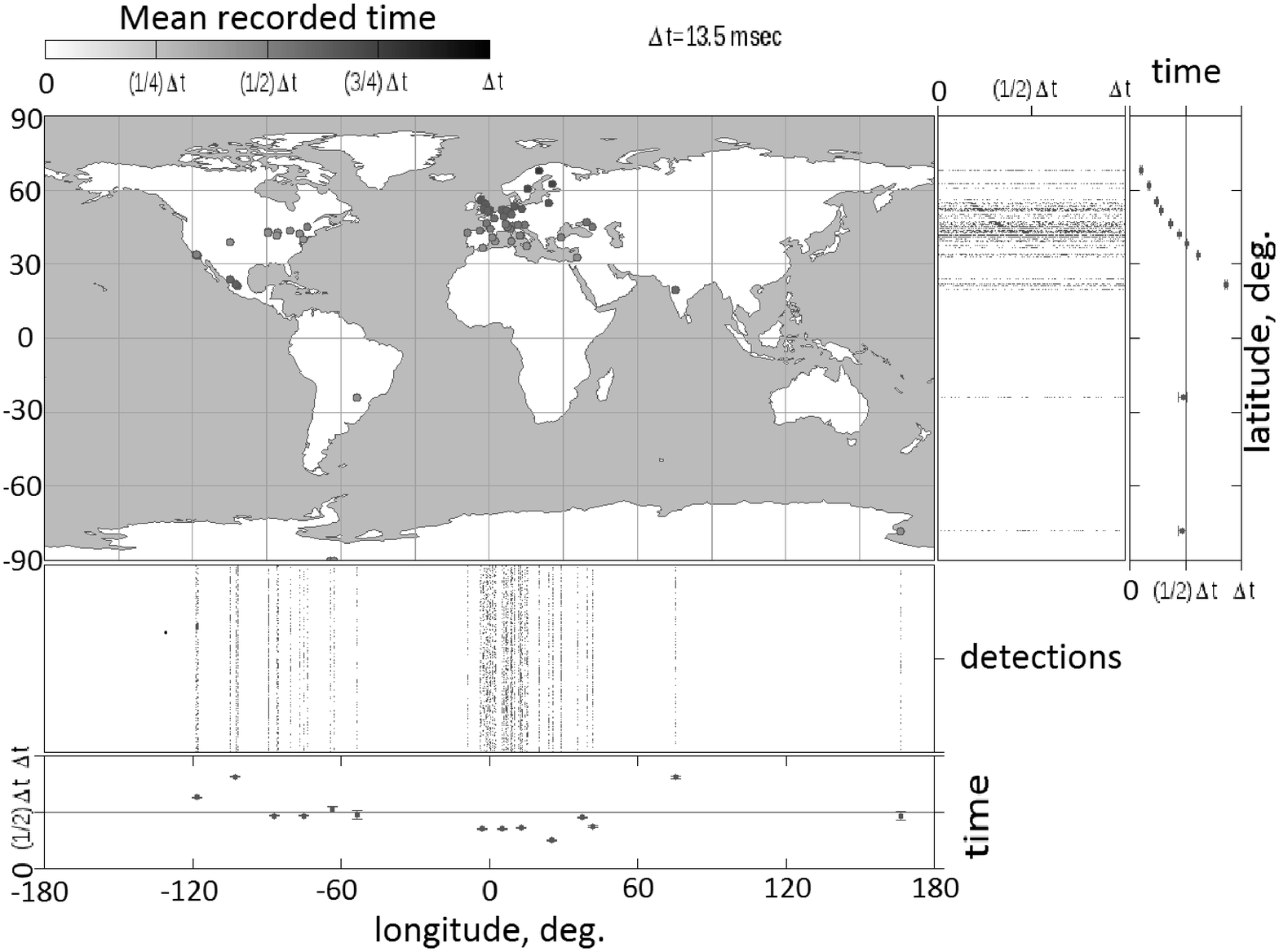,width = .85\linewidth}
\caption{SPS signal simulation.}\label{fig4}
\end{minipage}
\hfill
\begin{minipage}[t]{.45\linewidth}
\centering
\epsfig{file = 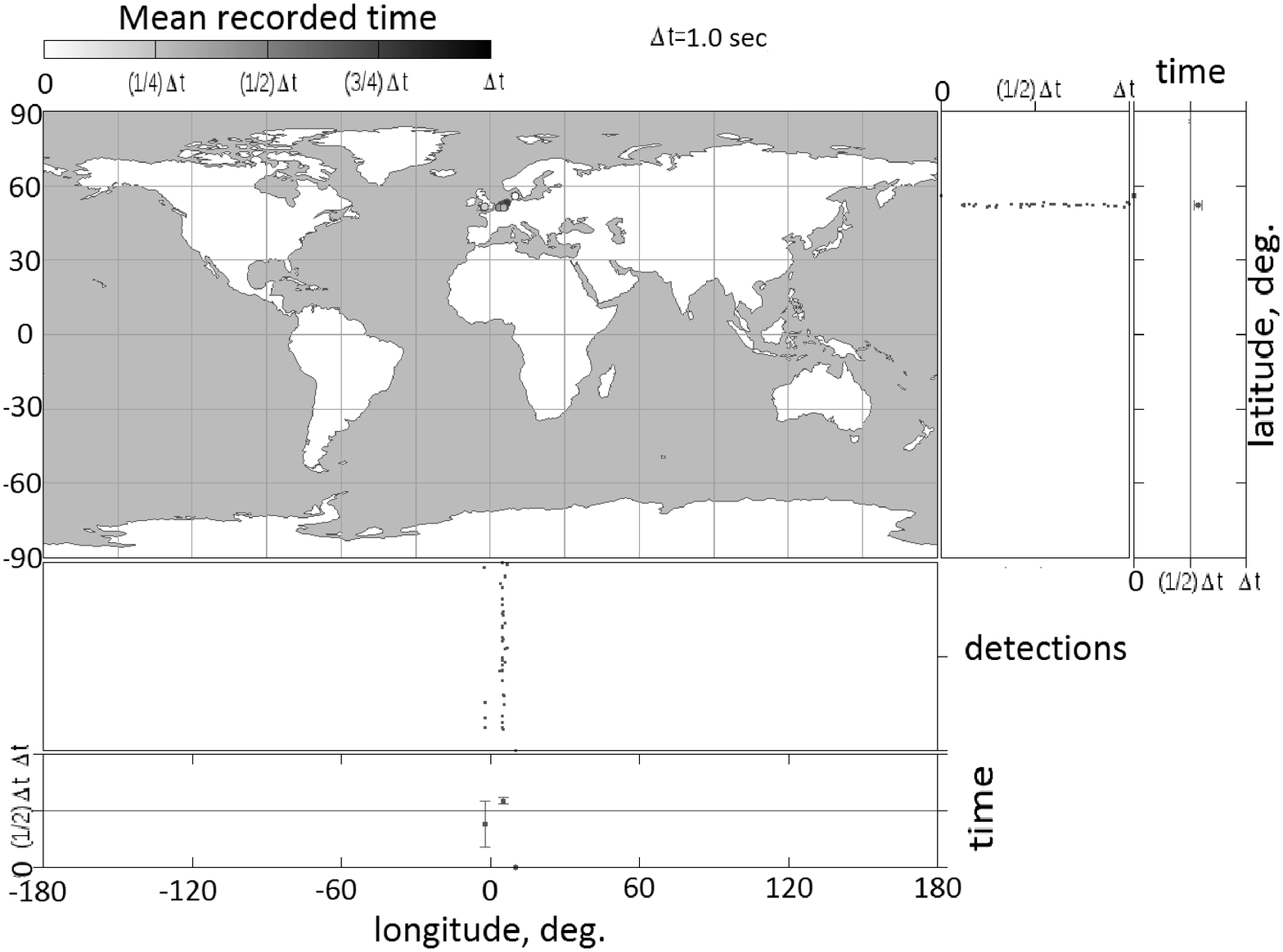,width = 0.85\linewidth}
\caption{An example of real data analysis.}\label{fig5}
\end{minipage}
\end{figure}

\indent \indent Within CREDO, a simple prototype monitoring system for automated data processing, with a current working name ``CREDO monitor'', has been already developed. It implements data analysis algorithms, mentioned above, and provides a simple visualization of the results. Data registered by every CREDO detector is stored in a central database server. In this regard, data from new cosmic ray detectors as well as from common electronic devices will be sent to the server in real-time or near real-time. Since CREDO is going to use available data gathered by other independently existing detector arrays in the same way, there will be periodic migration of relevant data from these systems for now. CREDO monitor follows several initially independent steps, later linked into a single script. Once data is available in the central server, it is ready to be processed by CREDO monitor. Data from all available individual detectors is first converted into a common format, then sorted in time and merged into the final form for further analysis and monitoring. Then the data is scanned to look for possible SPS signatures. As an example, we simulated the signal in search for a class ``C'' SPS. We look for a pattern composed of geographic locations of triggered detectors as a function of arrival times of particles in them (Fig.~\ref{fig4}). A scan performed in a small time window will give a flat line in the middle of the window for particle arrival times as a function of geographic coordinates, in the case when recorded signals arrive from uncorrelated, purely random particles. However, if recorded signals are from the considered class of SPS, we expect a departure from this flat line. Based on this concept, a citizen science interface to CREDO named ``Dark Universe Welcome'' \cite{DUW} has been developed, using which interested public, including scientists, students and science enthusiasts can participate in pattern recognition and classification of SPS events. Since there are no clear criteria on how the SPS signal being searched should exactly look like, no obvious ways of automated recognition are available so far. Therefore, it is very important to involve as many participants as possible for matching unique or ``strange'' features, which could be regarded as candidates for more detailed and thorough analysis. As it was mentioned previously, a number of detectors have already been integrated into the CREDO framework. Currently, data from available resources is periodically migrated to the data storage and computing center maintained at ACC Cyfronet AGH-UST \cite{CYFRONET}. An automated process is run every day for data migration, monitoring and analysis. The maps for visualization of real data analysis are also produced in the automated way (see Fig.~\ref{fig5}).

% as far as AASP has two-columns in its final version, try to fit your figures into the size of one column (width not more than 0.5\linewidth !) Pay attention to the size of axis titles: the fontsize should be readable after such ``fitting''.

\section*{\sc summary and outlook}
\indent \indent A common framework of existing detector arrays as well as individual detectors into a worldwide observatory 
provides us with an opportunity to probe unprecedented aspects of particle astrophysics and related areas. However, such
a framework was not in operation until recently. Our effort of integrating current cosmic-ray detecting infrastructure and facilities has already taken a formal shape and is in an ongoing progress. Once CREDO takes its final shape, it will open a new channel to explore the Universe, namely, the SPS channel. If observed, SPS can reveal yet unknown information on the interactions at energies close to the GUT scale, which in turn means an unprecedented opportunity to experimentally test dark matter models and scenarios, and probe interaction models and space-time properties in otherwise inaccessible energy range. On the other hand, non-observation of SPS will precise existing constraints on available theories and narrow down the area of ongoing searches for new physics. CREDO is aimed not only at fundamental physics questions, but can be helpful in a number of additional applications: alerting the astroparticle community on SPS candidates to enable a multi-channel data scan, potential research in interdisciplinary areas as geophysics and space-weather, integrating the scientific community (variety of science goals, detection techniques, wide cosmic-ray energy ranges etc.), helping non-scientists and young people to educate themselves and explore the Universe etc. Besides, with CREDO being developed and expanded, its participants can suggest their own ideas, which can lead to setting new goals and development of new projects. Everybody is welcome to join CREDO.

\section*{\sc acknowledgment}
\indent \indent This research has been supported in part by PLGrid Infrastructure. We warmly thank the staff at ACC Cyfronet AGH-UST, for their always helpful supercomputing support. The Dark Universe Welcome citizen science experiment was developed with the help of the ASTERICS Horizon2020 project. ASTERICS is a project supported by the European Commission Framework Programme Horizon 2020 Research and Innovation action under grant agreement n. 653477.

% read the instructions for the 

\end{document}